\documentclass[aps,prl,twocolumn]{revtex4-2}
\usepackage{amssymb, amsmath,  hyperref, graphicx, bm, color}
\begin{document}


\title{   
Quasi-Classical Gluon Fields and Low's Soft Theorem at Small $x$}


\author{Ming~Li}
\affiliation{Department of Physics, The Ohio State University, Columbus, OH 43210, USA}


\begin{abstract} 
In the high energy limit, soft gluons can be approximately described by quasi-classical gluon fields. It is well-known that  the gluon field is a pure gauge field on the transverse plane at eikonal order. We derived the complete next-to-eikonal order solutions of the classical Yang-Mills equations for soft gluons in the dense nuclear regime. Utilizing these solutions, it is shown that Low's soft theorem at small $x$ can be obtained by considering off-diagonal matrix elements of quasi-classical chromoelectric field between single gluon states in the dilute regime.  We further propose on extending Low’s soft theorem at small $x$ to incorporate the effects of gluon saturation in the dense regime.
\end{abstract}
\date{\today}
\maketitle
{\textbf{ Introduction.}} Using quasi-classical gluon fields to characterize soft gluons particularly in small $x$ physics has a long history \cite{Kovchegov:2012mbw}. In the high energy limit, nuclear objects are highly Lorentz contracted along the longitudinal direction so that a two-dimensional shockwave picture becomes applicable in describing high energy collisions \cite{Verlinde:1993te}. In the McLerran-Venugopalan (MV) model \cite{McLerran:1993ni, McLerran:1993ka} , soft gluon fields from large nuclei are solved from the classical Yang-Mills equations with color current sourced by hard gluons. This eikonal order quasi-classical gluon fields, valid in the parametrically dense regime $\mathcal{O}(1/g)$ in which nonlinear QCD interactions cannot be ignored \cite{Kovchegov:1996ty, Kovchegov:1997pc}, was used to estimate the Weiszacker-Williams gluon distribution inside a large nucleus in the quasi-classical approximation \cite{Jalilian-Marian:1996mkd}. Phenomenological applications in relativistic heavy-ion collisions \cite{Kovner:1995ja,Kovner:1995ts, Krasnitz:1999wc, Krasnitz:2000gz,Balitsky:2004rr,Schenke:2012wb, Gale:2012rq,Chen:2015wia} and high energy proton-nucleus collisions \cite{Schenke:2015aqa, Li:2021zmf, Li:2021yiv, Li:2021ntt} using the classical field approach were actively pursued in the past three decades \cite{Gelis:2010nm}
 .

Quasi-classical gluon field at eikonal order is insensitive to spin information of the nuclear object. To study spin related physical quantities particularly to understand the spin structure of proton at small $x$ \cite{Adamiak:2021ppq,Adamiak:2023yhz}, subeikonal order gluon field is needed. There have been a lot of efforts in recent years to identify the effective interactions at subeikonal order \cite{Laenen:2008gt, Jalilian-Marian:2017ttv, Chirilli:2018kkw,Kovchegov:2021iyc, Altinoluk:2021lvu, Li:2023tlw, Agostini:2023cvc} and to derive small $x$ evolution equations for polarized parton distributions inside proton/nucleus \cite{Kovchegov:2015pbl,Kovchegov:2016weo, Kovchegov:2017lsr, Kovchegov:2018znm, Chirilli:2021lif, Cougoulic:2022gbk}. Notably, the spin-dependent part of subeikonal order quasi-classical gluon fields was obtained using a diagrammatic approach in Lorenz gauge in \cite{Cougoulic:2020tbc}.

In this paper, we derived the complete solutions at subeikonal order for quasi-classical gluon fields  including both spin-dependent and spin-independent parts in the dense nuclear regime. The solutions are presented in the Lorenz gauge and can be readily transformed to the light-cone gauge. We found that, in addition to the piece found in  \cite{Cougoulic:2020tbc}, there exists a novel spin-dependent term induced by gluon saturation, which becomes significant only in the dense regime. The spin-independent part bears resemblance to the next-to-leading-order magnetic multipole expansion in two dimensions, as inferred from Ampere's law. Regarding the external color currents originating from hard gluons, our findings indicate that, apart from the color charge density at subeikonal order, the color spin density and transverse color current also serve as sources for the quasi-classical gluon fields.

\begin{figure}
    \includegraphics[width=0.35\textwidth]{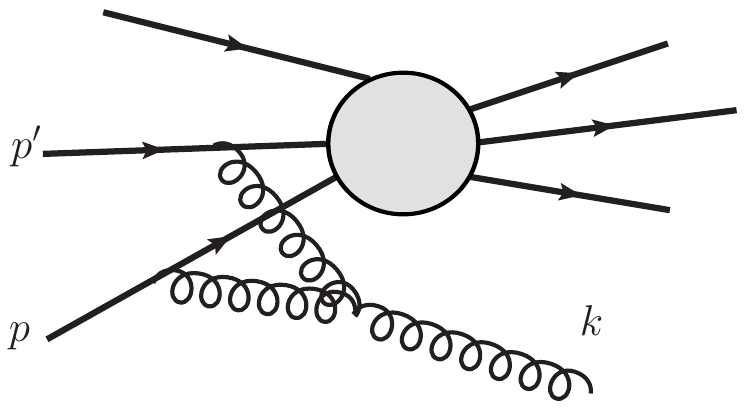}
    \caption{
		Schematic diagram showing gluon merging contribution to Low's soft theorem in the dense regime $\mathcal{O}(1/g)$.}
\label{fig:diag}		
\end{figure}

A closely related topic to soft gluons involves Low's soft theorem \cite{Low:1954kd, Gell-Mann:1954wra, Low:1958sn, Weinberg:1965nx, Burnett:1967km}, which states that at amplitude level, the radiative amplitude can be expressed as a factorized product of a soft factor and the nonradiative amplitude when taking the soft limit. In the past decades, there have been reviving interests in understanding Low's soft theorem in the Standard Model of particle physics and gravity from the perspective of asymptotic symmetries \cite{Strominger:2013lka,
Lysov:2014csa, Strominger:2017zoo}. Subleading order Low's soft theorem was rederived from various approaches \cite{Broedel:2014fsa, Bern:2014vva, White:2014qia, Larkoski:2014bxa}. 

It's worth noting that current research on Low's soft theorem has primarily focused on the dilute regime, where the soft gluon field is of order $\mathcal{O}(g)$. However, our ultimate objective is to understand Low's soft theorem in the dense regime, where the soft gluon field is  of order $\mathcal{O}(1/g)$. In this dense regime, nonlinear gluon merging processes, akin to those depicted in Fig.~\ref{fig:diag}, play an equally important role in contributing to Low's soft theorem, necessitating their resummation. The classical field approach provides a unique avenue for understanding Low's soft theorem in the dense regime, particularly in the small $x$ limit.  We propose that the small $x$ limit of Low's soft theorem can be obtained by calculating the off-diagonal matrix element of quasi-classical chromoelectric fields between incoming and outgoing nuclear states.  We explicitly demonstrate this proposal in the dilute regime by showing that the classical field approach effectively reproduces the small $x$ limit of Low’s soft theorem up to subleading order.

In the following, we first present the details of obtaining complete next-to-eikonal order solutions by solving classical Yang-Mills equations and then establish its equivalence to Low's soft theorem at small $x$.\\

%
{\textbf{ Eikonality Expansion of Yang-Mills Equations}}
We consider a pure glue theory and solve classical Yang-Mills equations. The inclusion of quarks and solving coupled Dirac equation for soft quark fields are left for a future work. 
Starting from the Yang-Mills action $S = -\frac{1}{4} \int_x F_{\mu\nu}^a F^{\mu\nu, a}$ with $F_{\mu\nu}^a = \partial_{\mu}A_{\nu}^a - \partial_{\nu}A_{\mu}^a + ig[A_{\mu}, A_{\nu}]^a$, one separates the full gluon field into soft gluon field $\mathcal{A}_{\mu}$ and hard gluon field $A_{\mu}$ according to their longitudinal momenta 
\begin{equation}
A_{\mu}^a \rightarrow \mathcal{A}_{\mu}^a + A_{\mu}^a.
\end{equation}
Let $\Lambda^+$ be some longitudinal momentum scale. Soft gluon fields represent modes with $k^+\ll \Lambda^+$ while hard gluon fields are associated with $k^+\gg \Lambda^+$. This division of degrees of freedom aligns with the spirit of Color Glass Condensate framework \cite{Iancu:2002xk} and the idea of rapidity factorization \cite{Balitsky:1995ub}. By implementing this separation and eliminating terms that are linear in hard gluon fields as well as terms exclusively concerning hard gluons, the effective action for soft gluons becomes 
\begin{equation}
\begin{split}
S = &-\frac{1}{4}\int_x\Big(\mathcal{F}^{\mu\nu}_a\mathcal{F}^a_{ \mu\nu} +4ig\mathcal{A}^{\mu}_a\left(\partial^{\nu}[A_{\mu}, A_{\nu}]^a  + [A_{\nu}, F^{\mu\nu}]^a\right)\\
&+2(ig)^2\Big([\mathcal{A}^{\mu}, \mathcal{A}^{\nu}]^a[A_{\mu}, A_{\nu}]^a +[\mathcal{A}_{\mu}, A_{\nu}]^a[\mathcal{A}^{\mu}, A^{\nu}]^a\\
& -[\mathcal{A}_{\mu}, A_{\nu}]^a [\mathcal{A}^{\nu}, A^{\mu}]^a)\Big).\\
\end{split}
\end{equation}
The equations of motion for soft gluons follow
\begin{equation}\label{eq:EOM}
\mathcal{D}_{\nu} \mathcal{F}^{\nu\mu} = J^{\mu}
\end{equation}
with the hard gluon color current
 \begin{equation}\label{eq:Jmu}
 J^{\mu} = ig[A_{\nu}, \bar{F}^{\mu\nu}] + ig\mathcal{D}^{\nu}[A^{\mu}, A_{\nu}].
 \end{equation}
 The covariant derivative is defined with respect to the soft gluon field $\mathcal{D}_{\nu} = \partial_{\nu} + ig[\mathcal{A}_{\nu}, \,\,]$. The field strength tensor for hard gluon field is defined correspondingly $\bar{F}^{\mu\nu} = \mathcal{D}^{\mu}A^{\nu} - \mathcal{D}^{\nu} A^{\mu} + ig[A^{\mu}, A^{\nu}]$. Our goal is to solve eq.~\eqref{eq:EOM} up to subeikonal order. 
 
 The general rule of counting the power of eikonality of gluon fields comes from their transformations under Lorentz boost \cite{Li:2023tlw}
 \begin{equation}
\begin{split}
&\mathcal{A}^+\longrightarrow \xi^{-1} \mathcal{A}^+ (\xi x^+, \xi^{-1} x^-, \mathbf{x}), \\
&\mathcal{A}^- \longrightarrow \xi \mathcal{A}^-(\xi x^+, \xi^{-1} x^-, \mathbf{x}),\\
&\mathcal{A}^i \longrightarrow \mathcal{A}^i(\xi x^+, \xi^{-1} x^-, \mathbf{x}).\\
\end{split}
\end{equation}
We used light-cone coordinates $x^{\pm} = (t\pm z)/\sqrt{2}$. Here $\xi = e^{-\omega}$ with $\omega$ characterizing the amount of  the Lorentz boost. In high energy limit, $\xi\rightarrow 0$. Expansion in eikonality is equivalent to Taylor expansion in powers of $\xi$ \cite{Bjorken:1970ah}.
\begin{equation}\label{eq:eikonality_expansion}
\begin{split}
&\mathcal{A}^+ = \xi^{-1} \mathcal{A}^+_{(-1)} + \mathcal{A}^+_{(0)} + \xi \mathcal{A}^+_{(1)} + \ldots,\\
&\mathcal{A}^- = \xi \mathcal{A}^-_{(1)}+\ldots ,\\
&\mathcal{A}^i = \mathcal{A}^i_{(0)} + \xi \mathcal{A}^i_{(1)} + \ldots .
\end{split}
\end{equation}
The arguments for the leading terms in the expansions are $(0^+, \tilde{x}^-, \mathbf{x}_{\perp})$, in which we have redefined $\tilde{x}^- = \xi^{-1} x^-$. 
 Similar eikonality expansions are understood for the color current $J^{\mu}$ and the field strength tensor $\mathcal{F}^{\mu\nu}$ when solving eq.~\eqref{eq:EOM} order by order in eikonality.\\

%
%
 
 {\textbf{Solutions in Lorenz Gauge}.} We derive solutions in the Lorenz gauge $\partial_{\mu} \mathcal{A}^{\mu} =0$. From the expansions in eq.~\eqref{eq:eikonality_expansion}, $\mathcal{A}^-=\mathcal{A}^i=0$ at the eikonal order. One looks for static solution $A^+_{(-1)}$ that satisfies 
\begin{equation}
\partial^2_{\perp} \mathcal{A}^+_{(-1)} = - J^+_{(-1)}.
\end{equation}
The solution at eikonal order is formally obtained as 
\begin{equation}
\mathcal{A}^+_{(-1)} = -\frac{1}{\partial_{\perp}^2} J^+_{(-1)} = \int d^2\mathbf{y}G(\mathbf{x}-\mathbf{y}) J^+_{(-1)}(x^-,\mathbf{y})
\end{equation}
with $G(\mathbf{x}-\mathbf{y}) = -\frac{1}{2\pi} \ln(|\mathbf{x}-\mathbf{y}|\Lambda)$ regularized by some IR scale $\Lambda$. 

At  subeikonal order, $\mathcal{A}^-=0$ and we look for solutions $\mathcal{A}^+_{(0)}\neq 0, \, \mathcal{A}^i_{(0)}\neq 0$ in the background of $\mathcal{A}^+_{(-1)}$. 
The Lorenz gauge condition reduces to $
\partial_{+}\mathcal{A}^+_{(0)} = -\partial_i \mathcal{A}^i_{(0)}
$. It is apparent that at subeikonal order $\mathcal{A}^+_{(0)}$ is dependent on the light-cone time $x^+$ while $\mathcal{A}^i_{(0)}$ is a static field.  Imposing these requirements, the Yang-Mills equations reduce to
\begin{equation}
\begin{split}
&2\partial_+\partial_-\mathcal{A}^+_{(0)} - \partial^2_{\perp} \mathcal{A}^+_{(0)} + 2ig\left[\mathcal{A}^+_{(-1)}, \partial_+ \mathcal{A}^+_{(0)}\right]\\
&\qquad\qquad\qquad\qquad\qquad+ 2ig\left[\mathcal{A}^i_{(0)}, \partial_i \mathcal{A}^+_{(-1)}\right] = J^+_{(0)},\\
&-\partial^2_{\perp} \mathcal{A}^i_{(0)} + ig\left[\mathcal{A}_{(0), j} , \partial^j \mathcal{A}_{(0)}^i\right] + ig\left[\mathcal{A}_{(0),j}, \mathcal{F}_{(0)}^{ji}\right] = J^i_{(0)}.
\end{split}
\end{equation} 
For the more general parametric regime in which $\mathcal{A}^{+}_{(-1)}, \mathcal{A}^+_{(0)}, \mathcal{A}^i_{(0)} \sim \mathcal{O}(1/g)$, the above equations are nonlinear and it is unclear whether closed-form analytic solutions exist or not. We consider subeikonal order solutions in the parametric regime   
$\mathcal{A}^+_{(0)}, \mathcal{A}^i_{(0)} \sim \mathcal{O}(1)$  as compared to the eikonal order field $\mathcal{A}^+_{(-1)}\sim \mathcal{O}(1/g)$. In this regime, the nonlinear equations are reduced to linear equations. This is also the parametric regime where helicity-dependent generalization of the McLerran-Venugopalan model was studied in \cite{Cougoulic:2020tbc}. The mixing between eikonality expansion and coupling constant expansion is a new feature at the subeikonal order. 

 With these considerations, the equations of motion to be solved are 
\begin{equation}
\begin{split}
& - \partial^2_{\perp} \mathcal{A}^+_{(0)}-2\partial_i \mathcal{D}_-A^i_{(0)}= J^+_{(0)},\\
&-\partial^2_{\perp} \mathcal{A}^i_{(0)}  = J^i_{(0)}.\\
\end{split}
\end{equation}
The covariant derivative here is defined as $\mathcal{D}_- = \partial_- + ig[\mathcal{A}^+_{(-1)}, \,\,\,]$.  The commutator part is the same order as that of the ordinary derivative because of $\mathcal{A}^+_{(-1)} \sim \mathcal{O}(1/g)$. 
We also consider $J^+_{(0)}$ to be of order $\mathcal{O}(1)$ and it is dependent on $x^+$. Current conservation at subeikonal order becomes $\partial_{+}J^+_{(0)}+ \partial_i J^i_{(0)}=0$. The solutions are readily obtained 
\begin{equation}\label{eq:A+Ai}
\begin{split}
&\mathcal{A}^+_{(0)} = - \frac{1}{\partial_{\perp}^2}\left(2\partial_i \mathcal{D}_-\mathcal{A}^i_{(0)} + J^+_{(0)}\right),\\
&\mathcal{A}^i_{(0)} = -\frac{1}{\partial^2_{\perp}}J^i_{(0)}.\\
\end{split}
\end{equation}
In subsequent sections, it will become evident that the transverse current $J^i_{(0)}$ contains a piece that is dependent on the helicity state of hard gluons. Therefore, in the dense regime, both the transverse field $\mathcal{A}^i_{(0)}$ and $A^+_{(0)}$,  which contains the commutator $ig[\partial_i A^+_{(-1)}, A^i_{(0)}]$, exhibit sensitivity to the helicity state of hard gluons. This latter helicity dependence arises solely from dense gluon effects and vanishes in the dilute regime. It is convenient to have the field strength tensor computed.
\begin{equation}\label{eq:Fmunu}
\begin{split}
&\mathcal{F}^{i+} = \left(\delta^{ij} - \frac{2\partial_i\partial_j}{\partial^2_{\perp}}\right)\mathcal{D}_- \frac{1}{\partial^2_{\perp}} J^j+ \frac{\partial_i}{\partial^2_{\perp}} J^+,\\
&\mathcal{F}^{ij} =- \frac{1}{\partial^2_{\perp}}(\partial^i J^j - \partial^j J^i),\\
&\mathcal{F}^{-+} =\frac{\partial_j}{\partial^2_{\perp}} J^j,\qquad \mathcal{F}^{-i}  = 0.\\
\end{split}
\end{equation}

%
%
{\textbf{ Low's Soft Theorem at Small $x$}.} 
In the dilute regime, Low's soft theorem at leading and subleading orders states that radiative amplitude can be expressed as soft factors acting on non-radiative amplitude when the radiated gluon becomes soft ($k\ll p_i$)
\begin{equation}
M^a(\{p_i\};  k) = \Big[ \mathcal{S}^{(-1)} + \mathcal{S}^{(0)}\Big] gT^a_{(i)}M(\{p_i\}) 
\end{equation}
The leading soft factor $\mathcal{S}^{(-1)}$ and the subleading soft factor $\mathcal{S}^{(0)}$ have the following expressions \cite{Broedel:2014fsa, Bern:2014vva, White:2014qia, Larkoski:2014bxa}\begin{equation}
\mathcal{S}^{(-1)} = \sum_{i}^n \frac{ p_i\cdot\varepsilon^{\ast}_{\lambda}(k)}{p_i \cdot k},\,\,\,\mathcal{S}^{(0)} = i\sum_i^n \frac{ \varepsilon^{\mu \ast}_{\lambda}(k)k^{\nu}J_{i,\mu\nu}}{p_i\cdot k}.
\end{equation}
It is considered that there are $n$ hard gluons with momenta $p_i$. $T^a_{(i)}$ is the color matrix associated with gluon ``$i$" and  $\varepsilon^{\mu}_{\lambda}(k)$ is the polarization vector of the soft gluon. $J_{\mu\nu} = L_{\mu\nu} + \Sigma_{\mu\nu}$ is the total angular momentum of the hard gluons.  The orbital angular momentum is $L_{\mu\nu} =i\left(p_{\mu} \frac{\partial}{\partial p^{\nu}} - p_{\nu} \frac{\partial}{\partial p^{\mu}}\right)$ and the spin operator for gluons is 
$(\Sigma^{\mu\nu})^{\alpha\beta} = i \left(g^{\mu\alpha}g^{\nu\beta} - g^{\mu\beta} g^{\nu\alpha}\right)$. Unlike the case for photon, the gluon soft theorem receives loop corrections \cite{Bern:2014oka, He:2014bga}. We restrict the discussion to tree level in this paper. 

Low's soft theorem up to subleading order was derived by  assuming $k^+, \mathbf{k} \ll p^+, \mathbf{p}$. On the other hand, in the small $x$ limit, one only requires that longitudinal momentum to be much smaller  $k^+\ll p^+$ while the transverse momentum is of the same order $\mathbf{k} \sim \mathbf{p}$. In the small $x$ limit, particularly into the gluon saturation regime, the typical transverse momentum of the soft gluons is characterized by the emerging gluon saturation scale $Q_s$. With $Q_s\gg \Lambda_{QCD}$, a perturbative treatment becomes applicable \cite{Kovchegov:2012mbw, Iancu:2002xk}.  To extract the small $x$ limit of Low's soft theorem, we expand the soft factors to linear order in power series expansion of $z = k^+/p^+\ll 1$. 

The leading order soft factor becomes
\begin{equation}\label{eq:S_leading_expansions}
\mathcal{S}^{(-1)} = -2\delta_{\sigma\sigma'}\left[\frac{\mathbf{k}^i }{\mathbf{k}^2} - z\left(\delta^{ij} - \frac{2\mathbf{k}^i\mathbf{k}^j}{\mathbf{k}^2}\right)\frac{\mathbf{p}^j}{\mathbf{k}^2} \right]\varepsilon^{i\ast}_{\lambda} + \mathcal{O}(z^2).\\
\end{equation}
Terms from expanding the soft factors are purely organized by counting the power of $z$. Although eq.~\eqref{eq:S_leading_expansions} should be understood with the assumption $\mathbf{k}\ll \mathbf{p}$, as will be shown in the next section, the validity of the first two terms actually extends to the momentum region $\mathbf{k} \sim \mathbf{p}$ in the limit $z\ll 1$.

For the subleading order soft factor, the part involving spin angular momentum becomes
\begin{equation}
\begin{split}
&\frac{i \varepsilon^{\ast}_{\lambda, \mu}(k) k_{\nu} (\Sigma^{\mu\nu})^{\alpha\beta} }{p\cdot k}\varepsilon_{\sigma}^{\alpha}(p)\varepsilon^{\beta,\ast}_{\sigma'}(p)\\
=&2z\sigma\delta_{\sigma\sigma'} \frac{i\epsilon^{ij} \varepsilon^{i \ast}_{\lambda}\mathbf{k}^{j} }{\mathbf{k}^2}+ \mathcal{O}(z^2).
\end{split}
\end{equation}
For the part involving orbital angular momentum, only derivatives with respect to the longitudinal momentum  $\partial/\partial p^+$ contribute. The nonvanishing components of the orbital angular momentum operator are $L_{-+} =  i p^+ \frac{\partial}{\partial p^+}$ and
$L_{i+} = i p_i \frac{\partial}{\partial p^+}$. One gets
\begin{equation}
\frac{i \varepsilon^{\mu\ast}_{\lambda}(k) k^{\nu} L_{\mu\nu}}{p\cdot k}\\
= 2\delta_{\sigma\sigma'}z\frac{ \mathbf{k}^i \varepsilon_{\lambda}^{i\ast}}{\mathbf{k}^2} \left( p^+ \frac{\partial}{\partial p^+}\right) +\mathcal{O}(z^2).
\end{equation}
It is interesting to observe that in the small $x$ limit the only components of angular momentum that contribute are $S^z=\frac{1}{2}\epsilon^{ij}\Sigma^{ij}$ for spin and $L^{+-}$ for orbital angular momentum. 

Putting all the pieces together, the soft factor up to subleading order in the small $x$ limit is 
\begin{equation}\label{eq:low_small_x}
\begin{split}
\mathcal{S} = & 2\delta_{\sigma\sigma'}\Big[-\frac{\mathbf{k}^i }{\mathbf{k}^2} + z\left(\delta^{ij} - \frac{2\mathbf{k}^i\mathbf{k}^j}{\mathbf{k}^2}\right)\frac{\mathbf{p}^j}{\mathbf{k}^2} \\
&+ z \sigma \frac{ i\epsilon^{ij}  \mathbf{k}^j}{\mathbf{k}^2} + z\frac{\mathbf{k}^i}{\mathbf{k}^2} \left(p^+\frac{\partial}{\partial p^+}\right)\Big]\varepsilon^{i\ast}_{\lambda} + \mathcal{O}(z^2).\\
\end{split}
\end{equation}

%
%

{\textbf{In the Dilute Regime}.} We show that eq.~\eqref{eq:low_small_x} can be obtained using the dilute limit of the quais-classical gluon field from eq.~\eqref{eq:Fmunu}
 by calculating its matrix element between incoming and outgoing single gluon states. 
In the dilute regime, the chromoelectric field $\mathcal{F}^{+i} \sim \mathcal{O}(g)$ and it has the expression
\begin{equation}\label{eq:Ai_dilute}
\mathcal{F}^{+i} = \frac{\partial^i}{\partial^2_{\perp}}J^+ - \left(\delta^{ij} - \frac{2\partial^i \partial^j}{\partial^2_{\perp}}\right) \frac{\partial_-}{\partial_{\perp}^2} J^j.
\end{equation} 
We use mode expansions for hard gluon fields to express the color current in terms of gluon creation and annihilation operators. For $J^+_a(k^+, \mathbf{x}) = \rho^a(\mathbf{x}) + J^+_{(0)}(k^+, \mathbf{x})$
\begin{equation}\label{eq:rhoa}
 \rho^a(\mathbf{x}) = igf^{abc} \int_{q^+} \hat{a}_{b, \lambda}(q^+, \mathbf{x}) \hat{a}^{\dagger}_{c, \lambda}(q^+, \mathbf{x}).
\end{equation}
and 
\begin{equation}
\begin{split}
J^{+}_{(0)}(k^+, \mathbf{x}) = & -igf^{abc} \frac{k^+}{2} \int_{q^+}\Big[ \hat{a}_{b, \lambda}(q^+, \mathbf{x})\frac{\partial}{\partial q^+}\hat{a}^{\dagger}_{c, \lambda}(q^+, \mathbf{x})\\
& - \frac{\partial}{\partial q^+}\hat{a}_{b, \lambda}(q^+, \mathbf{x}) \hat{a}^{\dagger}_{c, \lambda}(q^+, \mathbf{x})\Big]
\end{split}
\end{equation}
The total transverse current can be decomposed into
$J_a^i(\mathbf{x}) = -\epsilon^{il}\partial^l j_a(\mathbf{x}) +j_a^i(\mathbf{x})$
with the color spin density
\begin{equation}
j^a(\mathbf{x}) = igf^{abc}\int_{q^+} \frac{1}{q^+}\lambda \hat{a}_{b, \lambda}(q^+, \mathbf{x}) \hat{a}^{\dagger}_{c, \lambda}(q^+, \mathbf{x})
\end{equation}
and the spin-independent  transverse color current
\begin{equation}\label{eq:jla}
\begin{split}
j^l_{a}(\mathbf{x}) 
=&gf^{abc}\int_{q^+} \frac{1}{2q^+} \Big[\partial^{l}\hat{a}^{\dagger}_{c, \lambda}(q^+, \mathbf{x})\hat{a}_{b, \lambda}(q^+, \mathbf{x}) \\
&- \hat{a}^{\dagger}_{c, \lambda}(q^+, \mathbf{x})\partial^l\hat{a}_{b, \lambda}(q^+, \mathbf{x}) \Big].
\end{split}
\end{equation}
Unlike the case at eikonal order in which the effective degrees of freedom for hard gluons are completely determined by the color charge density $\rho^a(\mathbf{x})$,  one needs the color spin density $j^a(\mathbf{x})$, the transverse color current $j^i_a(\mathbf{x})$ and the subeikonal order correction $J^+_{(0)}(k^+, \mathbf{x})$ at subeikonal order.

Using these expressions, one calculates the matrix element of chromoelectric fields between incoming and outgoing single gluon Fock states
\begin{equation}\label{eq:F+i_matrix_element}
\begin{split}
&\left\langle 0 \left| \hat{a}_{d', \sigma'}(p^{\prime +}, \mathbf{p}')\mathcal{F}^{+i}_a(k^+, \mathbf{k})\hat{a}^{\dagger}_{d, \sigma}(p^+, \mathbf{p})\right|0\right\rangle\\
 =& gf^{add'} \delta_{\sigma\sigma'} 2\Big[  -\frac{\mathbf{k}^i}{\mathbf{k}^2} + z \frac{\mathbf{k}^i}{\mathbf{k}^2}\left(p^+\frac{\partial}{\partial p^+} \right) \\
 &\,\,+ \sigma z \frac{i\epsilon^{il}\mathbf{k}^l}{\mathbf{k}^2}  + z \left(\delta^{ij} - \frac{2\mathbf{k}^i\mathbf{k}^j}{\mathbf{k}^2}\right) \frac{\mathbf{p}^j}{\mathbf{k}^2} \Big]\\
 &\,\, \times (2\pi) 2p^+\delta(p^+-p^{\prime +}) (2\pi)^2\delta^{(2)}(\mathbf{p}'+\mathbf{k}-\mathbf{p}).\\  
 \end{split}
\end{equation}
Eq.~\eqref{eq:F+i_matrix_element} formally reproduces the soft factor at small $x$ given in eq.~\eqref{eq:low_small_x}. The overall factor $2$ accounts for the inclusion of both gluon emission and absorption in the classical fields, ensuring consistency with Low's soft theorem. Unlike eq.~\eqref{eq:low_small_x}, eq.\eqref{eq:F+i_matrix_element} is justified even when the transverse momenta are comparable $\mathbf{k} \sim \mathbf{p}$. Therefore, the small $x$ limit of Low's soft theorem turns out to be applicable when $\mathbf{k} \lesssim \mathbf{p}$, extending the conventional requirement $\mathbf{k} \ll \mathbf{p}$.

{\textbf{In the Dense Regime}.} In the dense regime, the chromoelectric field $\mathcal{F}^{+i} \sim \mathcal{O}(1/g)$ is given by eq.~\eqref{eq:Fmunu} rather than eq.~\eqref{eq:Ai_dilute}.   Low's soft factor at small $x$ is then calculated using incoming and outgoing states of dense nuclear objects such as large nuclei or protons at high energies,
\begin{equation}\label{eq:low_dense_regime}
\mathcal{S}_{\mathrm{Low}}\big|_{\mathrm{small}\, x} \sim  {}_H\langle P', s', a'| \mathcal{F}^{+i} | P, s, a\rangle_{H} 
\end{equation}
Here $|P, s, a\rangle_H$, which is also of order $\mathcal{O}(1/g)$,  represents the nuclear wavefunction for hard gluons with $P, s, a$ labeling the momentum, spin and color. The chromoelectric field $\mathcal{F}^{+i}$ for soft gluons is sourced by dense color currents of hard gluons eq.~\eqref{eq:Jmu},  $\mathcal{F}^{+i}$ can be derived by solving classical Yang-Mills equations in the dense regime, the nuclear wavefunction for hard gluons, which is presumably related to the hadronic/nuclear structure of proton/nucleus is in principle non-perturbative. Further analyses in the dense regime necessitate the modeling of off-diagonal elements of color currents beyond the conventional MV model.

%

{\textbf{Conclusions}.}  In this paper, we have proposed to study the small $x$ limit of Low's soft theorem using a quasi-classical field approach. To that end, we obtained the full subeikonal order solutions eq.~\eqref{eq:A+Ai} to the classical Yang-Mills equations with external currents in the dense regime. We explicitly demonstrate the equivalence between off-diagonal matrix elements of quasi-classical chromoelectric gluon field eq.~\eqref{eq:F+i_matrix_element} and Low’s soft theorem at small $x$ eq.~\eqref{eq:low_small_x} in the dilute regime. It is found that Low's soft theorem at small $x$ extends to the regime when the transverse momenta of  the soft gluons and the gluon emitting sources are comparable. Explicit analysis of Low's soft theorem towards dense regime using eq.~\eqref{eq:low_dense_regime} requires modeling of the nuclear wavefunction for hard gluons in addition to  the obtained dense fields $\mathcal{F}^{+i} \sim \mathcal{O}(1/g)$. In the current paper, Yang-Mills equations are solved in powers of $\xi$, with the Lorentz boost parameter $\xi\rightarrow 0$. 
Interestingly, the asymptotic symmetry approach to Low's soft theorem also involves solving Yang-Mills equations \cite{He:2015zea}. But it is carried out in the asymptotically flat space using power series expansion in terms of $1/r$ in the limit that the distance $r=\sqrt{x^2+y^2+z^2}$ approaches infinity $r\rightarrow \infty$ \cite{Strominger:2017zoo}. It would be interesting to study how these two approaches are connected, particularly addressing the topics of infrared safety of $S$-matrix \cite{He:2015zea} and color memory effects \cite{Pate:2017vwa, Ball:2018prg, Jokela:2019apz},  as well as unraveling the effects of dense gluons. \\


{\bf Acknowledgments.}
I thank  Alex Kovner and Yuri Kovchegov for reading the manuscript and valuable comments. I am also grateful to Yuri Kovchegov for very helpful and illuminating discussions.
The work is supported by the U.S. Department of Energy, Office of
Science, Office of Nuclear Physics under Award Number DE-SC0004286.


  \bibliography{softgluon}
\end{document}